\documentclass[sigconf,natbib=true]{acmart}

\usepackage{booktabs}
\usepackage{ragged2e}
\usepackage{nicematrix}
\usepackage{tabularx}
\usepackage{adjustbox}

\usepackage{graphicx}

\usepackage{enumitem}
\usepackage[linesnumbered,ruled]{algorithm2e}
\usepackage{colortbl}
\usepackage{multirow}
\usepackage{graphicx}
\usepackage[normalem]{ulem}
\useunder{\uline}{\ul}{}

\newcommand{\YYY}{\mathcal{Y}} %
\newcommand{\TTT}{\mathcal{T}} %
\newcommand{\SSS}{\mathcal{S}} %
\newcommand{\AAA}{\mathcal{A}} %

\newcommand{\LLL}{\mathcal{L}}

\definecolor{gblue}{RGB}{66,133,244}

\usepackage{arydshln}
\makeatletter
\def\adl@drawiv#1#2#3{%
        \hskip.5\tabcolsep
        \xleaders#3{#2.5\@tempdimb #1{1}#2.5\@tempdimb}%
                #2\z@ plus1fil minus1fil\relax
        \hskip.5\tabcolsep}
\newcommand{\cdashlinelr}[1]{%
  \noalign{\vskip\aboverulesep
           \global\let\@dashdrawstore\adl@draw
           \global\let\adl@draw\adl@drawiv}
  \cdashline{#1}
  \noalign{\global\let\adl@draw\@dashdrawstore
           \vskip\belowrulesep}}
\makeatother

\usepackage[normalem]{ulem} % gcm: 下划线:\sout{xxxx}

\usepackage{colortbl} % 引入colortbl包
\definecolor{mygray}{rgb}{0.9, 0.9, 0.9}
\definecolor{myred}{rgb}{0.68627451, 0.14117647, 0.09803922}
\newcommand{\myeq}[1]{{Eq.~(\ref*{eq:#1})}}
\newcommand{\mysec}[1]{{Section~\ref*{sec:#1}}}
\newcommand{\mytab}[1]{{Table~\ref*{tab:#1}}}
\newcommand{\myfig}[1]{{Fig.~\ref*{fig:#1}}}

\AtBeginDocument{%
  }

\copyrightyear{2025} 
\acmYear{2025} 
\setcopyright{acmlicensed}\acmConference[SIGIR '25]{Proceedings of the 48th
International ACM SIGIR Conference on Research and Development in
Information Retrieval}{July 13--18, 2025}{Padua, Italy}
\acmBooktitle{Proceedings of the 48th International ACM SIGIR Conference on
Research and Development in Information Retrieval (SIGIR '25), July 13--18,
2025, Padua, Italy}
\acmDOI{10.1145/3726302.3729981}
\acmISBN{979-8-4007-1592-1/2025/07}

\ccsdesc[500]{Information systems~Recommender systems}

\begin{document}

%%
%% The "title" command has an optional parameter,
%% allowing the author to define a "short title" to be used in page headers.

% \title{Modeling LLM-based Generative Recommendation as Generative Flow Networks}
% \title{Process-Supervised Language Model-based Recommendation Policy via Generative Flow Networks}

\title{Process-Supervised LLM Recommenders via Flow-guided Tuning} % for arxiv
% \title{Flower: Process-Supervised LLM-based Recommenders via Flow-guided Fine-Tuning} % for arxiv
% \title{Flow-guided Fine-Tuning for Diverse LLM-based Recommenders}

% 作者部分：显示在一排
% \author{Chongming Gao$^{1\#}$, Mengyao Gao$^{1\#}$, Chenxiao Fan$^{1}$, Shuai Yuan$^{2}$, Wentao Shi$^{1}$, Xiangnan He$^{1*}$}
% \author{
%   Chongming Gao\footnotemark, 
%   Mengyao Gao\footnotemark, 
%   Chenxiao Fan\textsuperscript{1}, 
%   Shuai Yuan\textsuperscript{2}, 
%   Wentao Shi\textsuperscript{1}, 
%   Xiangnan He\footnotemark
% }
% % 机构部分：列出机构和邮箱
% \affiliation{
%   \institution{
%   $^{1}$University of Science and Technology of China, Hefei, China\\
%   $^{2}$Hong Kong University of Science and Technology, Hong Kong, China}
%   \country{}
% }
% \email{chongminggao@ustc.edu.cn, mengyao0301@mail.ustc.edu.cn}
% \renewcommand{\thefootnote}{\fnsymbol{footnote}}
% \footnotetext{Both authors contributed equally to this research}
% \footnotetext{Corresponding author}

\author{Chongming Gao}
\authornote{Both authors contributed equally to this research.}
\email{chongminggao@ustc.edu.cn}
\affiliation{%
  \institution{University of Science and Technology of China}
  \city{Hefei}
  \country{China}
  % \country{}
}
\orcid{0000-0002-5187-9196}

\author{Mengyao Gao}
\email{mengyao0301@mail.ustc.edu.cn}
\orcid{0009-0004-5517-3563}
\authornotemark[1]
% \email{rjchen20@mail.ustc.edu.cn}
\affiliation{%
  \institution{University of Science and Technology of China}
  \city{Hefei}
  \country{China}
  % \country{}
}

\author{Chenxiao Fan}
\email{simonfan@mail.ustc.edu.cn}
\orcid{0009-0009-2509-7092}
\affiliation{%
  \institution{University of Science and Technology of China}
  \city{Hefei}
  \country{China}
  % \country{}
}

\author{Shuai Yuan}
\email{syuanaf@connect.ust.hk}
\orcid{0000-0001-6730-5755}
\affiliation{%
  \institution{Hong Kong University of Science and Technology}
  \city{Hong Kong}
  \country{China}
  % \country{}
}

\author{Wentao Shi}
\email{shiwentao123@mail.ustc.edu.cn}
\orcid{0000-0002-2616-6880}
\affiliation{%
  \institution{University of Science and Technology of China}
  \city{Hefei}
  \country{China}
  % \country{}
}

\author{Xiangnan He}
\authornote{Corresponding Author.}
\email{xiangnanhe@gmail.com}
\orcid{0000-0001-8472-7992}
\affiliation{%
  \institution{MoE Key Lab of BIPC, University of Science and Technology of China}
  \city{Hefei}
  \country{China}
  % \country{}
}

\renewcommand{\shortauthors}{Chongming Gao et al.}

%%
%% The abstract is a short summary of the work to be presented in the
%% article.
\begin{abstract}

While large language models (LLMs) are increasingly adapted for recommendation systems via supervised fine-tuning (SFT), this approach amplifies popularity bias due to its likelihood maximization objective, compromising recommendation diversity and fairness. 
To address this, we present \underline{\textbf{Flow}}-guided fin\underline{\textbf{e}}-tuning \underline{\textbf{r}}ecommender (Flower), which replaces SFT with a Generative Flow Network (GFlowNet) \cite{Bengio2024GFlowNet} framework that enacts process supervision through token-level reward propagation. Flower's key innovation lies in decomposing item-level rewards into constituent token rewards, enabling direct alignment between token generation probabilities and their reward signals. This mechanism achieves three critical advancements: (1) popularity bias mitigation and fairness enhancement through empirical distribution matching, (2) preservation of diversity through GFlowNet's proportional sampling, and (3) flexible integration of personalized preferences via adaptable token rewards. 
Experiments demonstrate Flower's superior distribution-fitting capability and its significant advantages over traditional SFT in terms of accuracy, fairness, and diversity, highlighting its potential to improve LLM-based recommendation systems.
The implementation is available via \url{https://github.com/Mr-Peach0301/Flower}.

\end{abstract}

%%
%% The code below is generated by the tool at http://dl.acm.org/ccs.cfm.
%% Please copy and paste the code instead of the example below.
%%
% \begin{CCSXML}
% <ccs2012>
%  <concept>
%   <concept_id>00000000.0000000.0000000</concept_id>
%   <concept_desc>Do Not Use This Code, Generate the Correct Terms for Your Paper</concept_desc>
%   <concept_significance>500</concept_significance>
%  </concept>
%  <concept>
%   <concept_id>00000000.00000000.00000000</concept_id>
%   <concept_desc>Do Not Use This Code, Generate the Correct Terms for Your Paper</concept_desc>
%   <concept_significance>300</concept_significance>
%  </concept>
%  <concept>
%   <concept_id>00000000.00000000.00000000</concept_id>
%   <concept_desc>Do Not Use This Code, Generate the Correct Terms for Your Paper</concept_desc>
%   <concept_significance>100</concept_significance>
%  </concept>
%  <concept>
%   <concept_id>00000000.00000000.00000000</concept_id>
%   <concept_desc>Do Not Use This Code, Generate the Correct Terms for Your Paper</concept_desc>
%   <concept_significance>100</concept_significance>
%  </concept>
% </ccs2012>
% \end{CCSXML}

% \ccsdesc[500]{Do Not Use This Code~Generate the Correct Terms for Your Paper}
% \ccsdesc[300]{Do Not Use This Code~Generate the Correct Terms for Your Paper}
% \ccsdesc{Do Not Use This Code~Generate the Correct Terms for Your Paper}
% \ccsdesc[100]{Do Not Use This Code~Generate the Correct Terms for Your Paper}

%%
%% Keywords. The author(s) should pick words that accurately describe
%% the work being presented. Separate the keywords with commas.
\keywords{Large Language Model Recommenders,Generative Flow Networks,\\
Diversity-aware Fine-tuning, Process Supervision
}
%% A "teaser" image appears between the author and affiliation
%% information and the body of the document, and typically spans the
%% page.

% \received{20 February 2007}
% \received[revised]{12 March 2009}
% \received[accepted]{5 June 2009}

%%
%% This command processes the author and affiliation and title
%% information and builds the first part of the formatted document.
\maketitle

\begin{figure}[!t]
\centering
\includegraphics[width=0.95\linewidth]{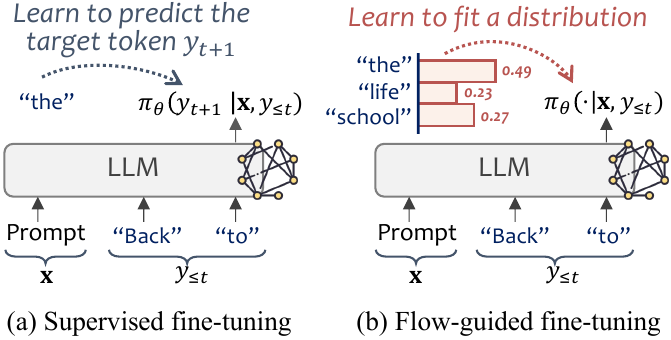}
% \vspace{-1mm}
\caption{Illustration of two tuning paradigms in LLM-based next-item recommendation tasks}
\label{fig:paradigms}
\end{figure}

\section{Introduction}

In recent years, recommendation systems powered by Large Language Models (LLMs) have emerged as a rapidly advancing field of research \cite{surveyLLM4rec,Lin2023survey}. Leveraging LLMs' vast repository of world knowledge and advanced learning capabilities, LLMs have demonstrated remarkable potential to enhance recommendation accuracy \cite{bao2024decoding,lin2024data}, improve fairness \cite{kddbias2024,jiang2024item}, and deliver more explainable \cite{yang2024fine,ma-etal-2024-xrec} and controllable \cite{chen2025dlcrec,lu-etal-2024-aligning,shen2024survey} recommendation results. These advancements contribute significantly to improving the overall user experience in recommendation systems.

While effective in general-purpose tasks, pretrained LLMs require supervised fine-tuning (SFT) to acquire the knowledge specific to downstream tasks. In recommendation tasks, preference modeling is a key objective. A commonly used SFT paradigm involves constructing instruction data from historical user behavior sequences, enabling the LLM to predict the next item based on prior interactions \cite{bao2023bi,bao2024decoding,gao2024sprec}. Items are typically represented by their titles, which consist of multiple tokens generated sequentially by the LLM. 
% During SFT, teacher forcing is applied to guide the model in predicting each token of the ground truth.
SFT leverages the cross-entropy (CE) loss to maximize the likelihood of the target labels. For example, in a next-movie recommendation task (\myfig{paradigms}(a)), the movie title ``Back to the Future'' is generated token by token. Given the prompt and the first two tokens, ``Back'' and ``to'', SFT optimizes the prediction of the next ground-truth token, ``the''.
While effective, SFT introduces two critical challenges for recommendation systems:
\begin{itemize}[leftmargin=*]
\item \textbf{Limited diversity.} SFT often drives models to produce high-probability tokens, leading to overfitting on dominant patterns in the training data \cite{li2024entropic}. As a result, the model generates homogeneous, less personalized recommendations–undermining the core goal of recommender systems.
\item \textbf{Popularity bias amplification.} SFT reinforces biases present in both fine-tuning data and pretraining corpora \cite{kddbias2024}, causing models to over-recommend popular items while underexposing niche content, which harms fairness and user experience.
\end{itemize}

To address these limitations, existing solutions in LLM-based recommendation systems (LRSs) can be broadly categorized into two approaches.  
The first approach involves modifying the SFT learning process. For example, assigning different weights to samples to adjust the learning loss \cite{jiang2024item} can mitigate category-specific popularity bias but fails to enhance the diversity of LLM-generated outputs. Another strategy is multi-stage SFT, where expert priors are integrated into successive stages to promote diversity \cite{chen2025dlcrec}. However, this method relies heavily on manually designed, multi-stage workflows, which are prone to error propagation and may exacerbate bias.
The second approach emphasizes post-SFT policy optimization to better align LLMs with human preferences, aiming to alleviate popularity bias and enhance diversity. For example, reinforcement learning with human feedback (RLHF) leverages reward signals to guide the recommender toward diverse outputs \cite{lu-etal-2024-aligning}. However, RLHF's reward maximization objective often leads to a collapse toward high-reward outcomes, which can further reduce diversity \cite{kirk2024understanding,yu2024flow}. More recently, Direct Preference Optimization (DPO) \cite{rafailov2024direct} has been applied to LRSs, using contrastive positive and negative samples to reduce popularity bias \cite{gao2024sprec,liao2024rosepo}. Despite its promise, DPO tends to induce distributional collapse by driving the probabilities of negative samples to zero \cite{IPO}, ultimately limiting recommendation diversity \cite{gao2024sprec}.

The shortcomings of these multi-stage pipelines or post-hoc corrections motivate our fundamental rethinking of alignment paradigms for LLM-based recommenders. Rather than patching SFT's limitations through auxiliary mechanisms, we propose \underline{\textbf{Flow}}-guided fin\underline{\textbf{e}}-tuning \underline{\textbf{r}}ecommender (Flower), which replaces conventional SFT with the generative flow network (GFlowNet)-based fine tuning. 
GFlowNet is a diversity-seeking reinforcement learning algorithm that trains policies to sample items with probabilities proportional to a given reward function rather than simply maximizing the reward \cite{Bengio2024GFlowNet}. 

By leveraging GFlowNet’s mechanism, we conceptualize the prefix tree of all feasible items as an irreversible flow network, where ``flow'' refers to the unnormalized probabilities moving from the root node to the leaf nodes. The reward assigned to each item corresponds to the flow at the leaf nodes. This framework enables us to compute the flow values at each branching point of the network, which serve as token-level rewards for next-token prediction. 
With these token-level rewards, Flower supervises the generation probabilities of each token during fine-tuning. As shown in \myfig{paradigms}(b), unlike SFT, which focuses on optimizing the prediction of a single ground truth token at a time, the objective of Flower is to align the policy (i.e., the predicted probability distribution over tokens) with the token-level reward distribution across all feasible tokens. 
Beyond fairness and diversity considerations, we further incorporate personalized user preferences into the derived token-level rewards using an auxiliary model, thereby enhancing both the accuracy and personalization of the recommendations.

Flower is a new fine-tuning paradigm superior to conventional SFT in enhancing diversity while maintaining accuracy. After tuning with Flower, additional alignment methods such as RLHF and DPO can still be applied. A comparison of different methods is provided in \mytab{intro}.
% Through extensive experiments, we first demonstrate Flower's superiority in fitting the training dataset. Furthermore, on three real-world sequential recommendation datasets, Flower outperforms SFT across accuracy, fairness, and diversity metrics. Finally, applying DPO or other alignment methods on top of the Flower-tuned model yields even better results compared to applying them on the original SFT model.
The main contributions are as follows:
\begin{itemize}[leftmargin=*]
\item We identify two key issues caused by the SFT mechanism in LRSs: limited diversity and popularity bias, both of which degrade recommendation performance.
\item We propose Flower, a fine-tuning paradigm that introduces the concept of flow to assign token-level rewards to all feasible next tokens, providing process-level supervision for LLMs.
\item Our token-level rewards are heuristically assigned, requiring no additional learning. This approach is simple yet effective and can be easily modified to incorporate personalized preferences.
\item Experiments validate the superiority of Flower over SFT in enhancing accuracy, diversity, and fairness. Moreover, these advantages are preserved even when further alignment methods such as PPO and DPO are applied.
\end{itemize}

\begin{table}[t]
% \small
\tabcolsep=2pt
\centering
\caption{Comparison of finetuning and alignment methods.}
\label{tab:intro}
\begin{tabular}{@{}lccm{4cm}@{}}
\toprule
\textbf{Paradigm} & \textbf{Method} & \textbf{Reward} & \multicolumn{1}{p{4cm}}{\centering \textbf{Objective}} \\
\midrule
\multirow{3}{*}{Finetuning} & SFT & / & \RaggedRight Maximize the probability for each ground-truth token. \\
\cdashlinelr{2-4}
& Flower & {\begin{tabular}[c]{@{}c@{}}Flow-guided \\ process\\ reward\end{tabular}} & \RaggedRight Align the next-token probability distribution with token-level reward distribution. \\
\midrule
\multirow{3}{*}{\begin{tabular}[c]{@{}c@{}}Preference\\ alignment\end{tabular}}  & RLHF & {\begin{tabular}[c]{@{}c@{}}Outcome\\ reward\end{tabular}} & \RaggedRight Maximize the reward of the generated item or item list. \\
\cdashlinelr{2-4}
& DPO & / & \RaggedRight Maximize the scores of the chosen item over the rejected. \\
\bottomrule
\end{tabular}
\label{tab:comparison}
\end{table}

\section{Preliminary}

This section introduces SFT for next-item recommendation with LLMs, followed by an overview of the basics of GFlowNets.

\subsection{SFT for Next-item Recommendation}
\label{sec:sft}

Next-item prediction is a fundamental task in recommender systems, where the goal is to predict the next item a user is likely to interact with, based on their historical behavior sequence. Leveraging the generative capabilities of LLMs, this task can be cast as generating the next recommended item directly.

Following the instruction-tuning paradigm of BIGRec \cite{bao2023bi}, we define the task as: given a prompt $\mathbf{x}$—typically describing the task and listing the user's previously interacted items—the model policy $\pi_\theta$ is trained to generate the next item $\mathbf{y}$, represented as a token sequence $\mathbf{y} = [y_1, y_2, \cdots, y_T]$.

The model is fine-tuned using standard cross-entropy loss with teacher forcing:
\begin{equation}\label{eq:sft}
\mathcal{L}_{\text{SFT}} = - \frac{1}{T} \sum_{t=1}^T \log \pi_\theta(y_t \mid \mathbf{x}, y_{1:t-1}),
\end{equation}
where $\pi_\theta$ denotes the model’s predicted probability for token $y_t$ given prior tokens and the input prompt.

After fine-tuning, the LLM is used to generate recommendations during inference. However, it may produce invalid or nonexistent items. To mitigate this, \citet{bao2023bi} propose a matching mechanism to align outputs with real items, while \citet{bao2024decoding} introduce constrained decoding to restrict token sampling to valid continuations. We adopt the latter to ensure recommendation validity.

\subsection{Problems in SFT-based Recommendations}
\label{sec:bias}

\begin{figure}[!t]
  \centering
  \includegraphics[width=\columnwidth]{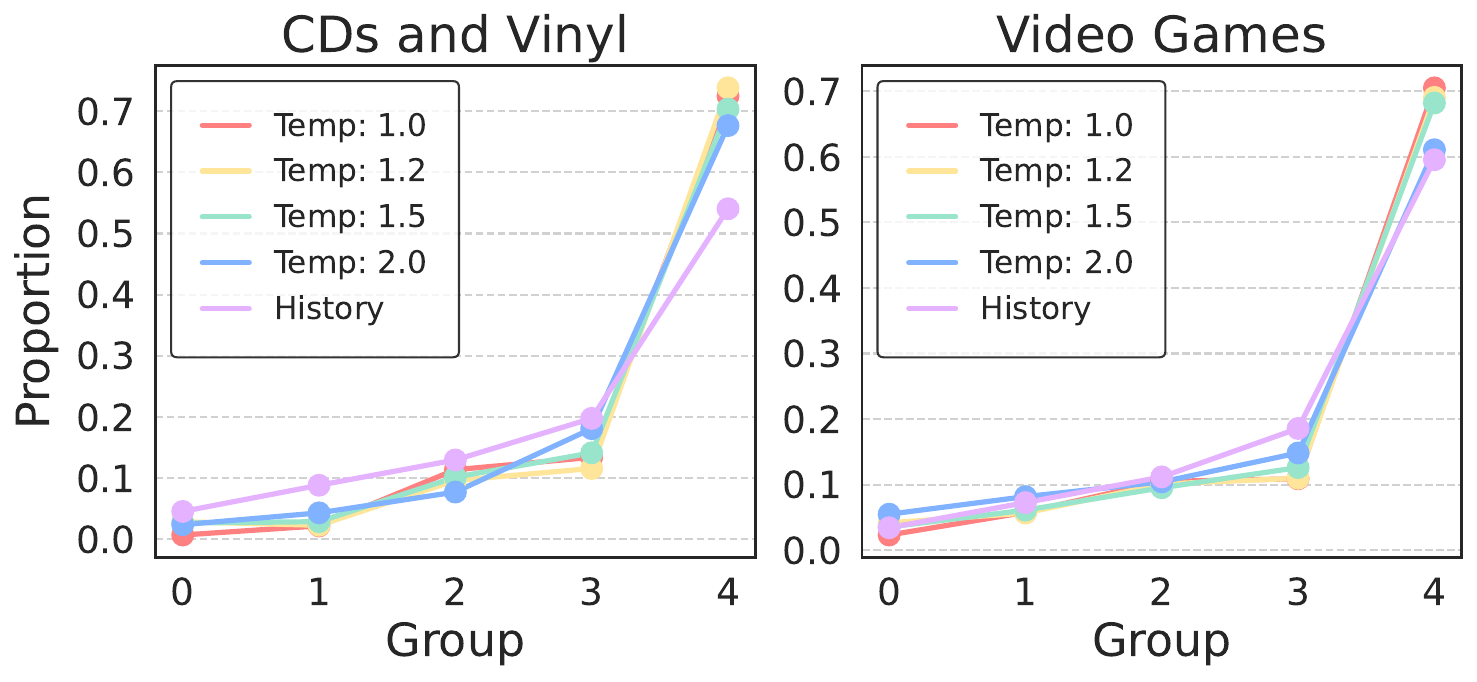} % 使用 \columnwidth 确保图片宽度适合单栏
  \caption{Distribution of the top-5 recommendations across five groups divided by their frequency of occurrence.} % 添加图片标题
  \label{fig:top5 bias} % 图片标签
\end{figure}

SFT introduces bias and fairness issues in LRSs \cite{kddbias2024,jiang2024item,gao2024sprec}. To illustrate this, we utilize the CDs and Video datasets from the Amazon Review datasets, dividing all items into five groups based on their frequency of occurrence in the historical sequences of the training data. After fine-tuning the LLM with BIGRec\footnote{Detailed experimental settings are provided in \mysec{experiment}}, we analyze the distribution of the top-5 recommended items across these five groups.
Furthermore, to investigate the influence of the temperature parameter during the inference stage, we conduct experiments with temperatures set to 1.0, 1.2, 1.5, and 2.0. 
% SFT introduces bias and fairness issues in LRSs \cite{kddbias2024,jiang2024item,gao2024sprec}. To illustrate this, we utilize the CDs and Video datasets from the Amazon Review datasets, dividing all items into five groups based on their frequency of occurrence in the historical sequences of the training data. After fine-tuning the LLM using BIGRec\footnote{The detailed experimental settings are provided in \mysec{experiment}}, we examine the distribution of the top-5 recommended items across these five groups.
% Additionally, to investigate the effect of the temperature parameter during the inference stage, we experiment with temperatures set to 1.0, 1.2, 1.5, and 2.0.

As shown in Figure \ref{fig:top5 bias}, Group 0 corresponds to the least popular items, while Group 4 represents the most popular ones. The top-5 recommendation results reveal a noticeable tendency to recommend items from the most popular group, disproportionately favoring when compared to the historical sequences. This highlights a clear presence of popularity bias, making the recommender system unfair to less popular items. 
Moreover, as the temperature increases, the unfairness is partially alleviated; however, both popularity bias and unfairness persist to varying degrees. 
% This suggests that while adjusting the temperature parameter can mitigate the issue to some extent, it does not fully resolve the inherent biases introduced during SFT.

% To analyze the bias of LLMs in recommending titles with varying levels of popularity, we compare the group distribution of top-5 recommendation results with that in the historical sequences of the training set on CDs and Video datasets in . As illustrated in Figure \ref{fig:top5 bias}, Group 0 represents the most unpopular group, while Group 4 represents the most popular one. The top 5 recommendation result exhibit a noticeable tendency to cluster towards the most popular group compared to the historical sequences, while the proportion of recommendations for titles in other groups is lower than their corresponding proportions in history. To investigate whether adjusting the temperature during recommendation generation could mitigate this bias, we set the temperature to 1.0, 1.2, 1.5, and 2.0, respectively. It can be observed that as the temperature increased, the unfairness was somewhat alleviated, but group-wise over-recommendation and under-recommendation still persisted.

% We adopt the method outlined by \citet{jiang2024item}, where $P(R)$, the target user history preference distribution, is approximated using the category distribution derived from offline training data. Furthermore, we also employ their proposed metrics, MGU@$K$, to systematically measure the degree of mismatch in our experiments.

\subsection{Generative Flow Networks (GFlowNets)}
\label{sec:GFlowNets}

Generative Flow Networks (GFlowNets) \cite{Bengio2021flow} are a class of generative models that learn stochastic policies for sequential decision-making. Inspired by physical flow systems, GFlowNets define unnormalized probabilistic flows to model diverse outcomes, allocating probability mass proportional to outcome rewards.

GFlowNets operate over a directed acyclic graph (DAG), where each path represents a sequence of actions from a root state to a terminal state. The flow through each path determines the probability of generating a particular outcome, enabling diverse and reward-aligned generation.

Formally, let $G = (\SSS, \AAA)$ denote the DAG, where nodes $s \in \SSS$ are \emph{states} and directed edges $(s_1 \to s_2) \in \AAA$ are \emph{actions}. If $(s_1 \to s_2)$ exists, then $s_2$ is a \emph{child} of $s_1$, and $s_1$ is its \emph{parent}. The graph has a unique \emph{initial state} $s_0$ with no parents, and a set of \emph{terminal states} $\YYY$ with no children. A \emph{trajectory} $\tau \in \TTT$ is a sequence of transitions $\tau = (s_m \to s_{m+1} \to \cdots \to s_n)$.

A \emph{forward policy} is a set of distributions $P_F(-|s)$ over the children of each nonterminal state $s \in \SSS$, inducing a trajectory distribution:
\begin{equation} \label{eq:forward_policy}
    P_F(\tau = (s_0 \to \cdots \to s_n)) = \prod_{i=0}^{n-1} P_F(s_{i+1} \mid s_i).
\end{equation}

% Given a nonnegative reward function $R : \YYY \to \mathbb{R}_{\geq 0}$, the GFlowNet objective is to learn $P_F$ such that the marginal probability of sampling $y \in \YYY$ is proportional to its reward:
% \begin{equation}\label{eq:reward_sampling}
%     R(y) = Z \sum_{\tau = (s_0 \to \cdots \to s_n = y)} P_F(\tau) \quad \forall y \in \YYY,
% \end{equation}
% where $Z = \sum_{y \in \YYY} R(y)$ is the normalization constant.

Given a nonnegative reward function $R : \YYY \to \mathbb{R}_{\geq 0}$, the objective of GFlowNets is to estimate a policy $P_F$ such that the likelihood of sampling $y \in \YYY$ is proportional to $R(y)$:
\begin{equation}\label{eq:reward_sampling}
    R(y) = Z \sum_{\tau = (s_0 \to \dots \to s_n = y)} P_F(\tau) \quad \forall y \in \YYY,
\end{equation}
where $Z$ is a normalization constant satisfying $Z = \sum_{y \in \YYY} R(y)$.

\begin{figure*}[!t]
    \centering
    \includegraphics[width=0.98\linewidth]{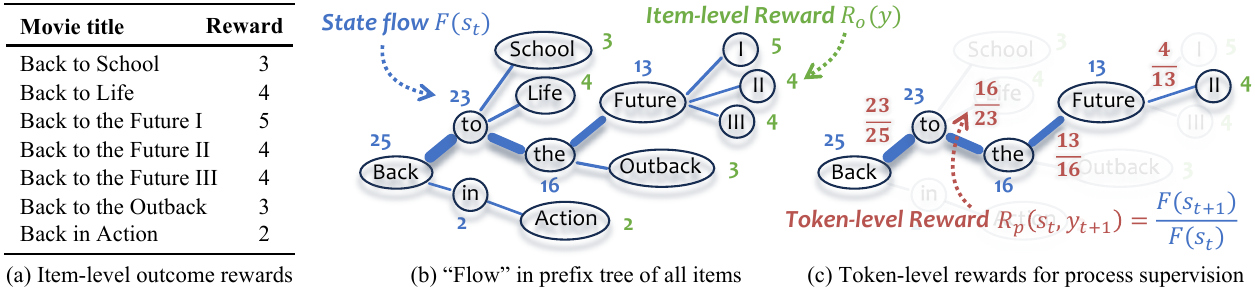}
    % \vspace{-3mm}
    \caption{Illustration of the prefix tree, state flow, item-level rewards, and flow-guided token-level rewards in Flower.}
    % \vspace{-4mm}
    % cumulative satisfaction, interaction sequence length, and the reward per turn in
    \label{fig:framework}
\end{figure*}

\subsection{Training Objective for GFlowNets}

% The sum over all $\tau \in \TTT$ in \myeq{reward_sampling} is intractable to compute directly, necessitating the use of auxiliary variables to facilitate learning the policy. In this work, we follow Subtrajectory Balance (SubTB) \cite{SubTB}, which use \emph{state flow} $F(s)$ as the auxiliary variable. A state flow $F(s)$ is a nonnegative scalar measuring the flow through state $s$, which is defined as:
The sum over all $\tau \in \TTT$ in \myeq{reward_sampling} is generally intractable, prompting the use of auxiliary variables to facilitate learning. We adopt the Subtrajectory Balance (SubTB) objective \cite{SubTB}, which introduces the \emph{state flow} $F(s)$ — a nonnegative scalar representing the total flow through state $s$:
\begin{equation}
    F(s):=F(\{\tau \in \mathcal{T}: s \in \tau\})=\sum_{\tau \in \mathcal{T}: s \in \tau} F(\tau).
\end{equation}
% These auxiliary variables represent different formulations of flow, such as the \emph{edge flow} $F(s \to t)$ used in Flow Matching \cite{Bengio2021flow}, \emph{state flow} $F(s)$ used in Detailed Balance (DB) \cite{Bengio2024GFlowNet,TB}, and \emph{initial state flow} $Z$ used in Trajectory Balance (TB) \cite{TB}. 
% In this work, we focus on the Subtrajectory Balance (SubTB) objective \cite{SubTB}, which generalizes the DB and TB conditions as special cases.
\medskip\noindent
\textbf{Subtrajectory Balance.}
% Let $F(s)$ be a state flow function. 
For any trajectory segment $\tau_{m:n} = (s_m \to \cdots \to s_n)$, the following constraint holds:
\begin{equation}\label{eq:subtb_constraint}
    F(s_m)\prod_{i=m}^{n-1}P_F(s_{i+1} | s_i; \theta)
    =
    F(s_n)\prod_{i=m}^{n-1}P_B(s_i | s_{i+1}; \theta),
\end{equation}
where $P_F(s_{i+1} | s_i; \theta)$ is the \emph{forward policy}, parameterized by $\theta$, representing the probability of taking action $(s_i \to s_{i+1})$ conditioned on the state $s_i$. Similarly, $P_B(s_i | s_{i+1}; \theta)$ is the \emph{backward policy}, modeling the reverse transitions. Here, $F(y) = R(y)$ if $y$ is terminal, and $F(s_0) = Z$ if $s_0$ is the initial state. 

% It is worth noting that the DB constraint \cite{Bengio2024GFlowNet} is a special case of \myeq{subtb_constraint} when the trajectory segment consists of a single action, i.e., $\tau = (s_m, s_{m+1})$. Likewise, the TB constraint \cite{TB} corresponds to the case where $\tau$ spans the entire trajectory, i.e., $\tau = (s_0, s_1, \cdots, y)$.

The Subtrajectory Balance constraint leads to the following \emph{subtrajectory balance objective}:
\begin{equation}\label{eq:subtb_objective}
    \ell_{\rm SubTB}(\tau_{m:n}) = 
    \left(
    \log \frac
    {F(s_m) \prod_{i=m}^{n-1} P_F(s_{i+1} | s_i; \theta)}
    {F(s_n) \prod_{i=m}^{n-1} P_B(s_i | s_{i+1}; \theta)}
    \right)^2.
\end{equation}
It is straightforward to observe that the objective in \myeq{subtb_objective} satisfies the desired condition of GFlowNets outlined in \myeq{reward_sampling}. 

\section{Method: Flower \includegraphics[width=0.03\textwidth]{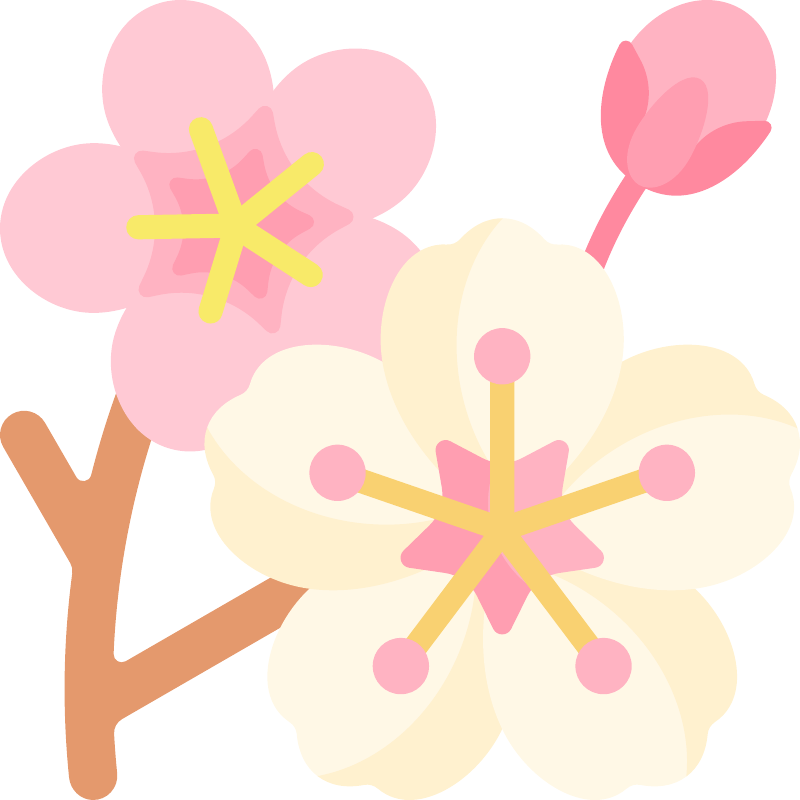}}
\label{sec:method}

In this section, we introduce the problem formulation and present how we fine-tune an LLM using our proposed method, Flower, to achieve process supervision.

\subsection{Problem Formulation}

% Let $\mathcal{I}$ denote the set of all valid items in the data set. Each item $i \in \mathcal{I}$ is represented by its title, a token sequence $\mathbf{y}^{(i)} = y_{1:T}^{(i)} = [y_1^{(i)}, y_2^{(i)}, \cdots, y_T^{(i)}]$, where $y_t^{(i)}$ is a token in the token space of the LLM. 
% The titles of all items collectively form a prefix tree, where items share common prefixes. For example, the movie titles ``Back to the Future II'' and ``Back to Life'' share the prefix ``Back to'' but diverge at subsequent tokens into two branches: ``the Future II'' and ``Life''. Note that in this example, words are used as tokens for illustration purposes. In practice, an item is automatically tokenized by an open-source LLM, such as LLaMA or Qwen. 

Let $\mathcal{I}$ denote the set of valid items in the dataset, where each item $i \in \mathcal{I}$ is represented by its title—a token sequence $\mathbf{y}^{(i)} = [y_1^{(i)}, y_2^{(i)}, \cdots, y_T^{(i)}]$, with each token $y_t^{(i)}$ belonging to the LLM's vocabulary. The titles collectively form a prefix tree, where shared prefixes define common paths. For instance, ``Back to the Future II'' and ``Back to Life'' share the prefix ``Back to'' before diverging. Note that in this example, words are used as tokens for illustration purposes. In practice, an item is automatically tokenized by an open-source LLM, such as LLaMA or Qwen. 

This prefix tree is a special DAG: nodes represent tokens, and edges denote valid transitions to the next token. The root corresponds to the empty sequence, and each path to a terminal node forms a complete item title.
We define: 
\begin{itemize}[leftmargin=*] 
\item \emph{State} $s_t$ as a prefix sequence $s_t=y_{\le t}=[y_0,\cdots,y_{t}]$, representing a node's path from the root. 
\item \emph{Action} $a_t$ as appending a valid token $y_{t+1}$, transitioning from the current state $s_t$ to a child state $s_{t+1}$.
\end{itemize}

\myfig{framework}(b) shows an example prefix tree for the 7 movie titles in \myfig{framework}(a). At the node ``to'', the state is ``Back to'', and valid actions lead to ``Back to School'', ``Back to Life'', or ``Back to the''.

The traversal aligns with the GFlowNets framework, where the probability of generating an item is the product of transition probabilities along its path. For example, the probability of ``Back to Life'' is computed as the product over ``Back'' $\to$ ``to'' $\to$ ``Life''.

% The prefix tree of all items forms a special DAG, where each node represents an exact token, and each directed edge corresponds to a possible next token. The root node represents the empty sequence, while paths from the root to terminal nodes represent complete item titles. 
% The state and action are defined as follows:
% \begin{itemize}[leftmargin=*]
%     \item The \emph{state} $s_t$ for a node $y_{t}$ is defined as a partial token sequence from the root node to this node, i.e., a prefix of one or more item titles: $s_t=y_{\le t}=[y_0,\cdots,y_{t}]$.
%     \item An \emph{action} $a_t$ is defined as the transition from the current state $s_t$ to a child state $s_{t+1}$ by appending a valid next token $y_{t+1}$ to the current sequence.
% \end{itemize}
% An example of this tree structure for all 7 movies in \myfig{framework}(a) is illustrated in \myfig{framework}(b). At the node ``to'', the state is ``Back to'', and three possible actions include transitioning to one of the following states: ``Back to School'', ``Back to Life'', or ``Back to the''.
% The flow through the prefix tree aligns with the GFlowNets framework, where the probability of generating a specific item is distributed along its corresponding path. For instance, the probability of generating the item ``Back to Life'' is computed as the product of probabilities along the path ``Back'' $\to$ ``to'' $\to$ ``Life''.

\medskip \noindent
\textbf{Remark:} 
This formulation casts generation as traversal through a prefix tree, ensuring only valid tokens are sampled at each step. It aligns model outputs with the dataset while leveraging GFlowNets to promote diversity and reward-proportional generation.

\subsection{LLM as The Policy for GFlowNets}

% \subsubsection{Policy}
We employ decoder-based LLMs to implement the policy in GFlowNets. 
Given the prompt $\mathbf{x}$, the model generates the tokens of an item $y$ sequentially: $y_1 \to \cdots \to y_T$. The probability of generating the item $y$ is defined as the product of the conditional probabilities of its tokens. Specifically, the forward policy \myeq{forward_policy} is expressed as:
\begin{equation}
    P_F(y) = \prod_{t=1}^{T} \pi_\theta(y_t | \mathbf{x}, y_{1:t-1}),
\end{equation}
where $y_{1:t-1}$ represents the sequence of tokens generated prior to step $t$, and $\pi_\theta$ is the decoder-based model parameterized by $\theta$. 
For example, the probability of generating the movie title ``Back to Life'' is calculated as:
\begin{equation}
\begin{aligned}
    P_F(\text{``Back to Life''}) 
    &= \pi_\theta(\text{``Back''} | \mathbf{x}) \cdot \pi_\theta(\text{``to''} | \mathbf{x}, \text{``Back''}) \\
    & \quad \cdot \pi_\theta(\text{``Life''} | \mathbf{x}, \text{``Back to''}).
\end{aligned}
\end{equation}

In the GFlowNets framework, the forward policy $P_F(s_{i+1} | s_i; \theta)$ in \myeq{subtb_constraint} is implemented as $\pi_\theta(y_t | \mathbf{x}, y_{1:t-1})$. Meanwhile, the backward policy $P_B(s_i | s_{i+1}; \theta)$ is always equal to 1, since each node in the prefix tree has only one parent.

\subsection{Flow-guided Token-level Reward}

Denote $R_o(y)$ as the \emph{outcome reward}, i.e., the item-level reward for generating an item $y$, we will derive the state flow $F(y)$ as below:

\subsubsection{State Flow on The Prefix Tree}
By setting the state flow at a terminal state equal to the outcome reward, i.e., $F(y) = R(y)$, and the initial flow equal to the total reward, $F(s_0) = Z = \sum_{y \in \mathcal{Y}} R_o(y)$, the flow for an intermediate state $s$ is recursively defined as the sum of the flows of its child states:
\begin{equation} \label{eq:flow}
    F(s) = \sum_{s' \in \text{Child}(s)} F(s').
\end{equation}
% An example is illustrated in \myfig{framework}(b), where the blue numbers near the nodes represent the state flows.

% \medskip
\noindent
\textbf{Remark:} Generally, estimating a proper flow in GFlowNets requires learning a flow function over the DAG \cite{Bengio2024GFlowNet}. However, benefiting from the prefix tree structure used in our approach, the flow can be directly computed without additional parameter estimation. This structure ensures computational simplicity and eliminates the need for a dedicated flow model.

Using the state flow, we can derive the objective of subtrajectory balance in \myeq{subtb_objective} for any trajectory $\tau_{m,n} = (s_m \to \cdots \to s_n)$ as:
\begin{equation}\label{eq:ourloss}
    \LLL(\tau_{m,n}) = 
    \left(
    \log \frac
    {F(s_m) \prod_{t=m}^{n-1} \pi_\theta(y_{t+1} | \mathbf{x}, y_{\le t})}
    {F(s_n)}
    \right)^2.
\end{equation}
% \medskip

\subsubsection{Process Reward}
To better illustrate how the policy is optimized through the flow mechanism, we define the \emph{process reward}, a token-level reward for generating the token $y_{t+1}$ given the prompt $\mathbf{x}$ and the previously generated tokens $y_{\le t}$. As shown in \myfig{framework}(c), the process reward is defined as:
\begin{equation}\label{eq:token_reward}
    R_p(s_t, a_t) = R_p(y_{\le t}, y_{t+1}) = \frac{F(s_{t+1})}{F(s_t)}.
\end{equation}
Using this definition, the objective \myeq{ourloss} can be rewritten as:
\begin{equation}\label{eq:ourloss_rewritten}
    % \mathcal{L}_{\rm SubTB}(\tau_{m,n}) = 
    \LLL_R(\tau_{m,n}) =\left(
    \sum_{t=m}^{n-1}\log 
    {\pi_\theta(y_{t+1} | \mathbf{x}, y_{\le t})} 
    -
    \sum_{t=m}^{n-1}\log
    {R_p(y_{\le t}, y_{t+1})}
    \right)^2.
\end{equation}
% \begin{equation}\label{eq:ourloss_rewritten}
%     % \mathcal{L}_{\rm SubTB}(\tau_{m,n}) = 
%     \sum_{t=m}^{n-1}\left(\log \frac
%     {\pi_\theta(y_{t+1} | \mathbf{x}, y_{\le t})}
%     {R_p(y_{\le t}, y_{t+1})}
%     \right)^2.
% \end{equation}
When the length of the subtrajectory is reduced to 2, the objective simplifies to directly fitting the policy $\pi_\theta(y_{t+1} | \mathbf{x}, y_{\le t})$ to the token-level reward $R_p(y_{\le t}, y_{t+1})$, thereby achieving process supervision.

\subsubsection{Reward Setting}
\label{sec:rewards}
To evaluate bias and fairness issues in LRSs, a common approach is to analyze the mismatch between the distribution of ground-truth user preferences and the distribution of model-predicted results \cite{kddbias2024}. Based on this perspective, we define the outcome reward $R_o(y)$ of an item $y$ as its frequency of occurrence in the historical sequences of the training data. 
The objective in \myeq{ourloss_rewritten} encourages the policy to generate items with a distribution aligned with the empirical data distribution, thereby addressing popularity bias and mitigating fairness issues. 

However, this reward remains static across all users and does not account for personalized preferences. To address this limitation, we introduce a \emph{preference score} $p_{ui}$, which predicts the likelihood of user $u$ liking item $i$. This score can be obtained from any auxiliary model, such as a traditional recommendation system. Given $p_{ui}$, we incorporate personalization by modifying the process reward term $\log R_p(y_{\le t}, y_{t+1})$ as:  
(1) $\frac{\log R_p(y_{\le t}, y_{t+1})}{p_{ui}}$, or  
(2) $\log (p_{ui} \cdot R_p(y_{\le t}, y_{t+1}))$.  
This adjustment effectively introduces user-specific information into the process rewards without altering the original flow derivation. In \mysec{RQ41}, we will compare the performance of these reward variants.

\medskip \noindent\textbf{Remark:}  
Many policy optimization methods employ complex process reward models (PRMs) \cite{lightman2024lets,wang2024math} to enhance reasoning in large language models \cite{setlur2025rewarding,zhang2024restmcts}, requiring significant computational resources for learning and verification. In contrast, our approach adopts heuristically assigned rewards, which are simple yet effective. This design avoids additional parameter learning and ensures efficient process supervision.

% This provides fine-grained guidance at the token level, ensuring that the generated sequences align closely with the desired flow and reward distributions.

% \medskip \noindent\textbf{Remark:}  
% There are numerous policy optimization techniques that utilize process reward models (PRMs) to enhance reasoning capabilities in large language models. These approaches typically require learning a complex PRM to verify and adjust the model's outputs, ensuring alignment with the desired objectives or reward structures. This often involves additional computational overhead and parameter estimation, which can make the training process more challenging and resource-intensive. 
% In contrast, our approach adopts a fundamentally different motivation. Instead of learning a complex PRM, the process rewards in our method are heuristically assigned. This design choice simplifies the reward computation while maintaining effectiveness, as the reward structure is directly derived from the prefix tree and aligned with the flow mechanism of GFlowNets. By leveraging this simple yet principled framework, we ensure that the model achieves process supervision without introducing unnecessary complexity or additional learning components. This simplicity not only reduces computational costs but also enhances the interpretability of the reward structure, making it more adaptable to practical applications.

\subsection{Fine-tuning LLMs through Process Rewards} 
To fine-tune the policy $\pi_\theta$, we integrate the original SFT loss $\mathcal{L}_{\rm{SFT}}$ from \myeq{sft} with the subtrajectory balance objective $\LLL_R$ in \myeq{ourloss_rewritten}. 
SFT is trained on offline datasets, while the subtrajectory balance objective is optimized on-policy by generating a batch of items, traversing their title set $\mathcal{T}$, and evaluating all feasible subtrajectories. The combined loss function of Flower is formulated as:
\begin{equation}\label{eq:flower_loss}
    \mathcal{L}_{\rm Flower}
    =
    \mathcal{L}_{\rm{SFT}} 
    +
    \lambda\sum_{\tau \in \mathcal{T}}\sum_{0 \leq m < n \leq T}\LLL_{R}(\tau_{m,n}),
\end{equation}
where $\tau_{m,n}$ represents a token subsequence from the title of a specific item. The hyperparameter $\lambda$ controls the trade-off between the SFT loss and the subtrajectory balance objective. 

This combined loss preserves the supervised performance of SFT while leveraging GFlowNets to promote diversity and reward-proportionality, enabling the policy to generate fairer and more representative items.

\section{Experiments}
\label{sec:experiment}
In this section, we conduct experiments to address the following research questions:

\begin{itemize}[leftmargin=*]
    \item \textbf{RQ1}: How effectively does Flower fit a specific distribution?
    \item \textbf{RQ2}: How do LLM-based methods perform in terms of accuracy, fairness, and diversity in the next-item recommendation task?  
    \item \textbf{RQ3}: What is the impact of using Flower as a reference policy on RL and DPO-based methods?
    \item \textbf{RQ4}: What are the effects of the key factors in Flower?  
    % \item \textbf{RQ4}: What impact do different reward settings have on the experimental results?
    % \item \textbf{RQ5}: How does the granularity of constraints influence the training process?
    % \item \textbf{RQ6}: What is the impact of $\lambda$ on recommendation performance?
\end{itemize}

\subsection{Experimental Setup} 

\begin{figure*}[!t]
  \centering
  \includegraphics[width=\textwidth]{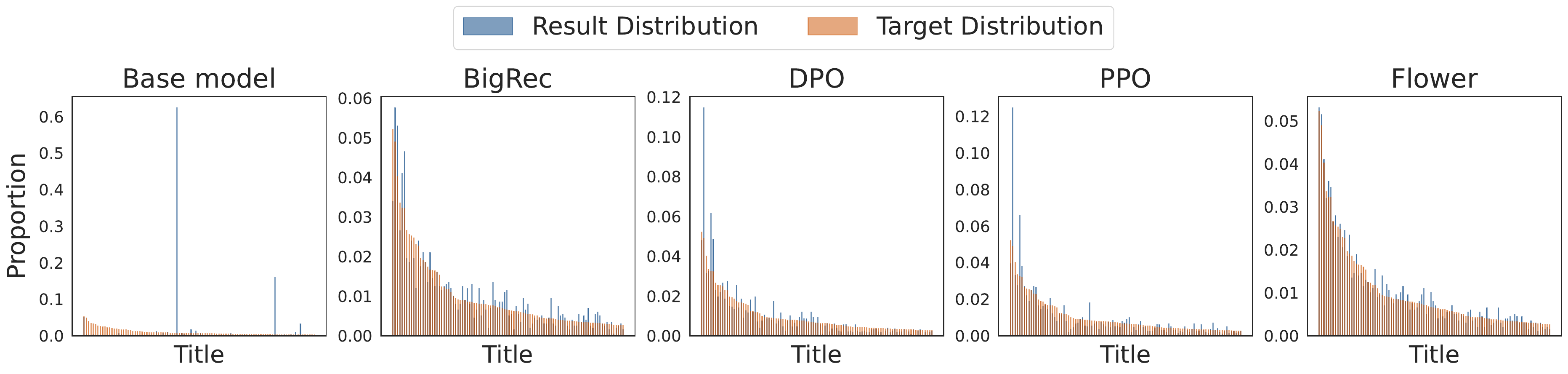} % 使用 \textwidth 确保图片宽度适合双栏
  % \caption{Comparison of distribution fitting performance on 100 movie titles.} % 添加图片标题
  \caption{Comparison of the distributions between the target set and the recommended results across 100 movie titles.}% 添加图片标题
  \label{fig:stage1 100} % 图片标签
\end{figure*}

\begin{table}[!t] 
\centering % 表格居中
\caption{Dataset statistics before and after processing.}
% \resizebox{\columnwidth}{!}{
\begin{tabular}{ccccc}
\toprule
\multicolumn{2}{c}{Dataset}                                                                                                       & \#User & \#Item & \#Interaction \\ \midrule
\multicolumn{1}{c}{\multirow{2}{*}{\begin{tabular}[c]{@{}c|@{}}CDs and Vinyl\\ 2015.10-2018.10\end{tabular}}} & Before & 440490 & 179277 & 815053        \\
\multicolumn{1}{c}{}                                                                                         & After  & 7685   & 5841   & 69249         \\ \midrule
\multicolumn{1}{c}{\multirow{2}{*}{\begin{tabular}[c]{@{}c|@{}}Video Games\\ 2015.10-2018.10\end{tabular}}}   & Before & 588656 & 45295  & 849496        \\
\multicolumn{1}{c}{}                                                                                         & After  & 9066   & 3858   & 70483         \\ \midrule
\multicolumn{1}{c}{\multirow{2}{*}{\begin{tabular}[c]{@{}c|@{}}Movies and TV\\ 2017.10-2018.10\end{tabular}}} & Before & 204439 & 53855  & 349292        \\
\multicolumn{1}{c}{}                                                                                         & After  & 3663   & 2653   & 31085         \\ \bottomrule
\end{tabular}
% }
\label{tab:Statistic Information} % 表格标签
\end{table}
\subsubsection{Dataset}
We conduct experiments on three real-world datasets from Amazon review data\footnote{https://cseweb.ucsd.edu/\textasciitilde jmcauley/datasets.html$\#$amazon$\_$reviews}, including CDs, Video Games, and Movies. These datasets contain user review data from May 1996 to October 2018. Following the preprocessing strategy in \cite{bao2023bi}, we truncate the dataset based on time information, remove unpopular users and items with fewer than five interactions, and limit the maximum item sequence length to 10. The datasets are chronologically split into training, validation, and test sets in an 8:1:1 ratio, detailed statistical information is provided in \mytab{Statistic Information}.

\subsubsection{Evaluation Protocal}
We evaluate top-k recommendation performance across three dimensions: Accuracy, Fairness, and Diversity. Accuracy and Fairness are assessed at the item level, while Diversity is evaluated at the word level.

For \textbf{Accuracy}, following prior work \cite{bao2023bi,bao2024decoding,jiang2024item}, we adopt two widely used metrics: Hit Ratio (HR@K) and Normalized Discounted Cumulative Gain (NDCG@K).  
\textbf{Fairness} is measured using DGU@K and MGU@K, which quantify the discrepancy between the group distribution of recommended titles in the top-k results and their distribution in the training set's historical sequences \cite{jiang2024item}. Titles are grouped by popularity: first, the frequency of each title in the training set is computed and sorted in descending order, then partitioned into eight equal-sized groups. Titles absent from the training set's historical sequences are assigned to the least popular group.  
To assess \textbf{Diversity}, we use two metrics: (1) Entropy of Words (H), which calculates the entropy of all English words in the recommended titles, and (2) Type-Token Ratio (TTR), defined as the ratio of unique words to the total number of words in the recommendations.

In the comparison results, we report NDCG@5, HR@5, DGU@10, MGU@10, H, and TTR as evaluation metrics.

\subsubsection{Baseline}
We select one traditional sequential recommendation model and several SFT-based LRSs as baselines:
\begin{itemize}[leftmargin=8pt]
    \item \textbf{SASRec} \cite{kang2018self} is a widely used sequential recommendation baseline employing a self-attention mechanism.
    \item \textbf{BIGRec} \cite{bao2023bi} is one of the earliest and most classic supervised fine-tuning (SFT) methods for directly generating item titles.
    \item \textbf{Temp} \cite{bao2024decoding} adjusts the temperature coefficient for inference based on the trained BIGRec model.
    \item $\boldsymbol{D^3}$ \cite{bao2024decoding} enhances BIGRec by removing length normalization to address amplification bias and incorporating a text-free assistant model, SASRec, in the inference stage to mitigate the homogeneity issue and improve recommendation diversity. 
    \item \textbf{IFairLRS} \cite{jiang2024item} improves fairness in BIGRec by balancing recommendations across categories through weighting the SFT loss.
\end{itemize}

\subsubsection{Implementation Details}
For SASRec, we optimize using binary cross-entropy loss and the Adam optimizer, with a learning rate in [1e-2, 1e-3, 1e-4], a batch size of 1024, and weight decay in [1e-3, 1e-4, 1e-5, 1e-6]. For LLM-based methods, we use Qwen2.5-1.5B-Instruct as the base model, with the learning rate set to 3e-4, batch size to 128, maximum training epochs to 7, early stopping patience to 2, and optimize models using the AdamW optimizer. For experiments related to the temperature coefficient, we adjust it within the range of [1.2, 1.5, 2.0]. Considering the scale difference between the flow-guided loss and the SFT loss, $\lambda$ is in the range of [0.01, 0.005, 0.001, 0.0005, 0.0001]. Unless otherwise specified, we use the process reward $\frac{\log R_p}{p_{ui}}$ described in \mysec{rewards} for Flower, where the preference score $p_{ui}$ is obtained from SASRec.

\begin{table}[!t]
\tabcolsep=4.1pt
\renewcommand\arraystretch{1.0}
% \caption{Comparison of distribution fitting performance on 1500 movie titles. Kullback-Leibler (KL) divergence and Jensen-Shannon (JS) Divergence are calculated on title and token level. For KL divergence, T and R represent target distribution and test result distribution, respectively.}
\caption{
Quantitative results of the distribution mismatch between the target set (T) and the recommended results (R) across 1500 movie titles. KL divergence and JS divergence are computed at both the title and token levels.}
\begin{tabular}{lccccc}
\toprule
               & Base Model                     & BIGRec                        & DPO                            & PPO                           & Flower                                 \\ \midrule
Title KL(T||R) & \cellcolor[HTML]{FFFFFF}20.235 & \cellcolor[HTML]{FFFFFF}5.114 & \cellcolor[HTML]{FFFFFF}17.765 & \cellcolor[HTML]{FFFFFF}9.985 & \cellcolor[HTML]{FFFFFF}\textbf{0.961} \\
Title KL(R||T) & \cellcolor[HTML]{FFFFFF}4.117  & \cellcolor[HTML]{FFFFFF}0.788 & \cellcolor[HTML]{FFFFFF}2.703  & \cellcolor[HTML]{FFFFFF}1.466 & \cellcolor[HTML]{FFFFFF}\textbf{0.190} \\
Title JS       & \cellcolor[HTML]{FFFFFF}0.513  & \cellcolor[HTML]{FFFFFF}0.184 & \cellcolor[HTML]{FFFFFF}0.449  & \cellcolor[HTML]{FFFFFF}0.341 & \cellcolor[HTML]{FFFFFF}\textbf{0.047} \\
Token KL(T||R) & \cellcolor[HTML]{FFFFFF}16.999 & \cellcolor[HTML]{FFFFFF}3.646 & \cellcolor[HTML]{FFFFFF}13.653 & \cellcolor[HTML]{FFFFFF}6.549 & \cellcolor[HTML]{FFFFFF}\textbf{1.982} \\
Token KL(R||T) & \cellcolor[HTML]{FFFFFF}5.291  & \cellcolor[HTML]{FFFFFF}1.307 & \cellcolor[HTML]{FFFFFF}2.940  & \cellcolor[HTML]{FFFFFF}2.174 & \cellcolor[HTML]{FFFFFF}\textbf{0.838} \\
Token JS       & \cellcolor[HTML]{FFFFFF}0.565  & \cellcolor[HTML]{FFFFFF}0.291 & \cellcolor[HTML]{FFFFFF}0.458  & \cellcolor[HTML]{FFFFFF}0.429 & \cellcolor[HTML]{FFFFFF}\textbf{0.217} \\ \bottomrule
\end{tabular}
\label{tab:stage1 1500}
\end{table}

\begin{table*}[]
\small
\tabcolsep=1.8pt
\renewcommand\arraystretch{1.0}
\caption{Performance of all methods evaluated in terms of accuracy, fairness, and diversity. The best results are bolded.}
\begin{tabular}{c|cccccc|cccccc|cccccc}
\hline
         & \multicolumn{6}{c|}{{\color[HTML]{000000} CDs and Vinyl}}                                             & \multicolumn{6}{c|}{{\color[HTML]{000000} Video Games}}                                               & \multicolumn{6}{c}{{\color[HTML]{000000} Movies and TV}}                                              \\ \hline
         & \textbf{NDCG}$\uparrow$   & \textbf{HR}$\uparrow$     & \textbf{DGU}$\downarrow$   & \textbf{MGU}$\downarrow$   & \textbf{H}$\uparrow$     & \textbf{TTR}$\uparrow$   & \textbf{NDCG}$\uparrow$   & \textbf{HR}$\uparrow$     & \textbf{DGU}$\downarrow$   & \textbf{MGU}$\downarrow$   & \textbf{H}$\uparrow$     & \textbf{TTR}$\uparrow$   & \textbf{NDCG}$\uparrow$   & \textbf{HR}$\uparrow$     & \textbf{DGU}$\downarrow$   & \textbf{MGU}$\downarrow$   & \textbf{H}$\uparrow$     & \textbf{TTR}$\uparrow$   \\ \hline
SASRec   & 0.0641          & 0.0851          & 0.184          & 0.038          & 9.188          & 0.124          & 0.0369          & 0.0544          & 0.167          & 0.033          & 8.229          & 0.050          & 0.0902          & 0.1072          & 0.138          & 0.032          & 8.892          & 0.167          \\
BIGRec   & 0.0573          & 0.0715          & 0.217          & 0.045          & 5.900          & 0.006          & 0.0326          & 0.0466          & 0.151          & 0.029          & 7.504          & 0.004          & 0.0930          & 0.1134          & 0.123          & 0.028          & 8.297          & 0.018          \\
Temp     & 0.0503          & 0.0627          & 0.222          & 0.044          & 6.202          & 0.006          & 0.0306          & 0.0444          & 0.129          & 0.026          & 7.307          & 0.004          & 0.0852          & 0.1061          & 0.139          & 0.027          & 8.145          & 0.018          \\
D3       & \textbf{0.0812} & \textbf{0.0999} & 0.355          & 0.072          & 7.635          & 0.013          & 0.0413          & 0.0607          & 0.220          & 0.041          & 7.645          & 0.005          & \textbf{0.1007} & \textbf{0.1225} & 0.147          & 0.033          & 8.348          & 0.020          \\
IFairLRS & 0.0621          & 0.0762          & 0.217          & 0.045          & 6.420          & 0.007          & 0.0396          & 0.0568          & 0.144          & 0.030          & 7.699          & 0.005          & 0.0957          & 0.1170          & 0.159          & 0.043          & 8.048          & 0.015          \\
\rowcolor[HTML]{EFEFEF} 
Flower   & 0.0700          & 0.0885          & \textbf{0.075} & \textbf{0.021} & \textbf{7.919} & \textbf{0.013} & \textbf{0.0543} & \textbf{0.0799} & \textbf{0.108} & \textbf{0.023} & \textbf{7.750} & \textbf{0.005} & 0.0959          & 0.1199          & \textbf{0.076} & \textbf{0.026} & \textbf{8.808} & \textbf{0.023} \\ \hline
\end{tabular}
\label{tab:main result}
\end{table*}

\begin{table*}[!t]
%\footnotesize
\small
\tabcolsep=1.8pt
\renewcommand\arraystretch{1.0}
\caption{Performance comparison of Flower (F) and BIGRec (B) as reference policies of RL and DPO-based methods.}
\begin{tabular}{c|cccccc|cccccc|cccccc}
\hline
              & \multicolumn{6}{c|}{CDs and Vinyl}                                                                     & \multicolumn{6}{c|}{Video Games}                                                                       & \multicolumn{6}{c}{Movies and TV}                                                                     \\ \hline
              & \textbf{NDCG}$\uparrow$   & \textbf{HR}$\uparrow$     & \textbf{DGU}$\downarrow$   & \textbf{MGU}$\downarrow$   & \textbf{H}$\uparrow$     & \textbf{TTR}$\uparrow$   & \textbf{NDCG}$\uparrow$   & \textbf{HR}$\uparrow$     & \textbf{DGU}$\downarrow$   & \textbf{MGU}$\downarrow$   & \textbf{H}$\uparrow$     & \textbf{TTR}$\uparrow$   & \textbf{NDCG}$\uparrow$   & \textbf{HR}$\uparrow$     & \textbf{DGU}$\downarrow$   & \textbf{MGU}$\downarrow$   & \textbf{H}$\uparrow$     & \textbf{TTR}$\uparrow$   \\ \hline
B\_PPO    & 0.0519          & 0.0640          & 0.246          & 0.049          & 5.670          & 0.005          & 0.0282          & 0.0401          & 0.191          & 0.035          & 7.204          & 0.004          & 0.0871          & 0.1075          & 0.175          & 0.033          & 8.114          & 0.016          \\
B\_S-DPO   & 0.0712          & 0.0908          & 0.104          & 0.025          & 8.539          & 0.016          & \textbf{0.0671} & \textbf{0.0900} & 0.083          & 0.020          & 8.223          & 0.008          & 0.1037          & 0.1232          & \textbf{0.070} & 0.022          & 9.068          & 0.025          \\
B\_RosePO & 0.0641          & 0.0810          & 0.105          & 0.023          & \textbf{8.627} & 0.017          & 0.0599          & 0.0786          & 0.286          & 0.057          & \textbf{8.546} & 0.008          & 0.1012          & 0.1178          & 0.145          & 0.030          & 9.347          & 0.027          \\
B\_DMPO   & 0.0718          & 0.0890          & 0.083          & 0.016          & 8.275          & 0.015          & 0.0424          & 0.0622          & 0.056          & 0.015          & 8.254          & 0.007          & 0.0960          & 0.1199          & 0.076          & 0.026          & 8.807          & 0.023          \\
\rowcolor[HTML]{EFEFEF}
F\_PPO    & 0.0620          & 0.0788          & 0.085          & 0.023          & 7.574          & 0.011          & 0.0565          & 0.0757          & 0.124          & 0.024          & 7.561          & 0.005          & 0.0963          & 0.1196          & 0.083          & 0.028          & 8.751          & 0.022          \\
\rowcolor[HTML]{EFEFEF}
F\_S-DPO   & \textbf{0.0772} & \textbf{0.0944} & 0.085          & 0.019          & 8.326          & 0.016          & 0.0636          & 0.0834          & 0.075          & 0.016          & 8.393          & 0.007          & \textbf{0.1042} & \textbf{0.1269} & 0.073          & \textbf{0.017} & 9.159          & 0.026          \\
\rowcolor[HTML]{EFEFEF}
F\_RosePO & 0.0701          & 0.0872          & 0.127          & 0.028          & 8.608          & 0.017          & 0.0608          & 0.0799          & 0.305          & 0.059          & 8.501          & \textbf{0.008} & 0.1012          & 0.1214          & 0.188          & 0.037          & \textbf{9.361} & \textbf{0.028} \\
\rowcolor[HTML]{EFEFEF}
F\_DMPO   & 0.0731          & 0.0913          & \textbf{0.063} & \textbf{0.012} & 8.545          & \textbf{0.017} & 0.0644          & 0.0869          & \textbf{0.043} & \textbf{0.013} & 8.233          & 0.007          & 0.0974          & 0.1211          & 0.072          & 0.022          & 8.721          & 0.024          \\ \hline
\end{tabular}
\label{tab:reference policy}
\end{table*}

% \subsection{Method Advantage Verification (RQ1)}  
% \subsection{Unbiased Fitting Capability (RQ1)}  
\subsection{Distribution Fitting Capability (RQ1)}  
\label{sec:RQ1}

Before applying our method to personalized recommendation problems, we first evaluate the distribution fitting capabilities of Flower compared to SFT, DPO, and PPO in a history-free recommendation scenario using the Movies and TV dataset. In this setup, the LLM is prompted to recommend a movie without providing any history or examples, aiming to assess how well each method aligns with the distribution of the training set, i.e., the target item set.
We use the same target item set for all methods. After tuning, each method generates recommendations equal in size to the tuning dataset. We evaluate the mismatch between the distribution of items in the target set and the generated recommendations.

\subsubsection{Qualitative Visualization}
For illustration, we create the target set by sampling 100 items from the Movie dataset, preserving their interaction frequencies as shown in \mytab{Statistic Information}. We employed Qwen2.5-1.5B-Instruct as the base model. BIGRec uses all interactions in the target set as training data. For DPO, we randomly select one item as the chosen response and another less-interacted item as the rejected response. For PPO, we assign normalized interaction counts as item-level rewards. Additionally, we report the recommendation results of the base model (pre-trained LLM without tuning) as a reference.

The item-level distributions are illustrated in \myfig{stage1 100}. The base model's recommendations concentrate on a few specific titles, while BIGRec skews heavily toward popular titles, a bias further amplified in DPO and PPO. In contrast, Flower effectively learns the target distribution, capturing titles with varying popularity and mitigating the unfairness observed in other methods.

\subsubsection{Quantitative Analysis}
To quantify these observations, we increase the target set to 1500 items and employ Qwen2.5-3B-Instruct as the base model. We compute the Kullback-Leibler (KL) divergence and Jensen-Shannon (JS) divergence between the generated and target distributions at both the token and item levels. As shown in \mytab{stage1 1500}, Flower achieves superior distribution fitting compared to other methods, validating its ability to enhance diversity and align with the target distribution.

\begin{table*}[!t]
\caption{Recommendation performance under different reward settings. The best results are highlighted in bold.}
\vspace{-1mm}
\small
\tabcolsep=1.3pt
\renewcommand\arraystretch{1.0}
\begin{tabular}{c|cccccc|cccccc|cccccc}
\hline
\multicolumn{1}{l|}{} & \multicolumn{6}{c|}{CDs and Vinyl}                                                                  & \multicolumn{6}{c|}{Video Games}                                                                     & \multicolumn{6}{c}{Movies and TV}                                                                   \\ \hline
\multicolumn{1}{l|}{} & \textbf{NDCG}$\uparrow$   & \textbf{HR}$\uparrow$     & \textbf{DGU}$\downarrow$   & \textbf{MGU}$\downarrow$   & \textbf{H$\uparrow$}     & \textbf{TTR}$\uparrow$   & \textbf{NDCG}$\uparrow$   & \textbf{HR}$\uparrow$     & \textbf{DGU}$\downarrow$   & \textbf{MGU}$\downarrow$   & \textbf{H}$\uparrow$     & \textbf{TTR}$\uparrow$   & \textbf{NDCG}$\uparrow$   & \textbf{HR}$\uparrow$     & \textbf{DGU}$\downarrow$   & \textbf{MGU}$\downarrow$   & \textbf{H}$\uparrow$     & \textbf{TTR}$\uparrow$   \\ \hline
$\log R_p$                  & \textbf{0.0712} & 0.0880          & 0.071          & 0.019          & 7.885          & 0.013          & 0.0372          & 0.0539          & 0.102          & 0.020          & 7.651          & 0.005          & 0.0930          & 0.1123          & 0.084          & \textbf{0.022} & \textbf{8.944} & \textbf{0.025} \\
$\log(R_p \cdot p_{ui})$             & 0.0700          & \textbf{0.0898} & \textbf{0.060} & \textbf{0.018} & 7.902          & 0.013          & 0.0366          & 0.0532          & \textbf{0.092} & \textbf{0.018} & 7.595          & 0.005          & 0.0905          & 0.1156          & 0.115          & 0.026          & 8.540          & 0.019          \\
$\frac{\log R_p}{p_{ui}}$                & 0.0700          & 0.0885          & 0.075          & 0.021          & \textbf{7.919} & \textbf{0.013} & \textbf{0.0543} & \textbf{0.0799} & 0.108          & 0.023          & \textbf{7.750} & \textbf{0.005} & \textbf{0.0953} & \textbf{0.1192} & \textbf{0.076} & 0.027          & 8.808          & 0.023          \\ \hline
\end{tabular}
\label{tab:reward comparision}
\end{table*}

% \subsection{Main Results (RQ2)}  
% \subsection{Performance in Next-item Recommendation (RQ2)}  
\subsection{Next-item Recommendation Results (RQ2)}  
\label{sec:RQ2}

We evaluate the recommendation performance of Flower and baseline methods on the next-item recommendation task using three open-world datasets. Unlike traditional recommendation models that rely on item IDs to represent titles, LLM-based models are inherently constrained by the text generation nature of LLMs, leading to a diversity gap compared to traditional methods. As a result, we exclude SASRec from diversity comparisons. The overall experimental results are summarized in \mytab{main result}, and the key observations are as follows:
\begin{itemize}[leftmargin=*]
    \item Compared to baseline methods, Flower achieves optimal fairness and diversity across all datasets. While Flower's accuracy is second only to $D^3$ on the Video Games dataset, it demonstrates consistent advantages in fairness and diversity. Notably, $D^3$ improves diversity by integrating SASRec collaborative information during inference but suffers from the poorest fairness among all methods, with a significant gap compared to Flower. This highlights Flower's balanced performance across all metrics.
    \item For methods specifically aimed at enhancing fairness, Flower outperforms IFairLRS across all metrics and datasets. This indicates that, compared to IFairLRS's approach of reweighting the entire title, Flower's use of probabilistic process supervision at the token level through GFlowNets achieves better results. 
    % By breaking down the title-level reward into token-level guidance, Flower ensures fine-grained supervision, resulting in more accurate and fair recommendations with greater diversity.
    \item Temp improves fairness and diversity over BIGRec by introducing higher randomness during inference. However, this improvement comes at the expense of accuracy. In contrast, Flower's approach of decomposing title rewards into token-level conditional probabilities allows for strict token-wise supervision. This mechanism ensures a balance between randomness and control, leading to simultaneous improvements in accuracy, fairness, and diversity without sacrificing any single dimension.
    \item Both Flower and $D^3$ leverage collaborative information from SASRec. However, $D^3$ focuses solely on accuracy optimization, neglecting fairness and diversity by not imposing additional constraints. As a result, $D^3$ achieves higher accuracy but suffers from significant fairness degradation. Conversely, Flower integrates fairness considerations into the collaborative information, enabling simultaneous improvements in accuracy, fairness, and diversity. This showcases Flower's ability to balance multiple objectives effectively.
\end{itemize}

% \begin{itemize}[leftmargin=*]
%     \item Compared to baseline methods, Flower achieves optimal fairness and diversity across all datasets. Except on the Video Games dataset, Flower's accuracy is second only to $D^3$. However, with the introduction of SASRec collaborative information during inference, $D^3$ exhibits improved diversity but demonstrates the poorest fairness among all methods, with a significant gap compared to Flower.
%     \item For methods aimed at enhancing fairness, Flower outperforms IFairLRS on all metrics and datasets. This indicates that, compared to reweighting the entire title, imposing probabilistic process supervision on each token through GFlowNets, yields (请扩写) better training outcomes.
%     \item Temp improves fairness and diversity over BIGRec but sacrifices accuracy due to increased inference randomness. In contrast, Flower, by decomposing title rewards into token-level conditional probabilities, achieves strict token-wise supervision, leading to improvements in accuracy, fairness, and diversity.
%     \item Both Flower and $D^3$ incorporate collaborative information from SASRec, but $D^3$ does not consider fairness or impose additional constraints, resulting in accuracy optimization at the cost of fairness degradation. In contrast, Flower integrates fairness information with collaborative information, achieving improvements across three dimensions simultaneously.
% \end{itemize}

% \subsection{Work as Reference Policy (RQ3)}  
% \subsection{Potential as Reference Policy (RQ3)} 
\subsection{Flower as a Reference Policy (RQ3)}  

\label{sec:RQ3}  
Existing RLHF and DPO-based methods are typically fine-tuned on top of SFT. In contrast, Flower is a fine-tuning framework designed to address the diversity and unfairness issues inherent in SFT. We investigate the performance of Flower and BIGRec (an SFT-based method) as reference policies for preference alignment methods, such as RL and DPO-based approaches.

We select three DPO-based recommendation methods for comparison:
\textbf{DMPO} \cite{DMPO} is framework that bridges the gap between LLMs and recommendation tasks by sampling multiple negative items as rejected responses.
\textbf{S-DPO} \cite{chen2024softmax} is method that incorporates multiple negative samples in user preference data and generalizes pairwise DPO loss to a softmax ranking loss.
\textbf{RosePO} \cite{liao2024rosepo} is a general framework that combines negative sampling strategies with personalized uncertainty to improve fairness, unbiasedness, and robustness. In our experiments, we sample rejected items based on item popularity distributions for comparison with RosePO.
For RL-based methods, we implement PPO \cite{schulman2017proximal}, using the frequency of each title in the training set as the reward.

As shown in \mytab{reference policy}, similar to how Flower outperforms BIGRec, preference alignment methods based on Flower generally achieve better performance compared to their BIGRec-based counterparts. This demonstrates the potential of Flower-tuned methods for downstream applications. 

Notably, RosePO excessively suppresses popular items, leading Flower's results—initially closer to the target distribution than BIGRec's—to deviate further from the target distribution compared to B\_RosePO, resulting in poorer fairness for F\_RosePO. Furthermore, due to the simplicity of its reward design, PPO fails to effectively optimize performance. However, Flower-based F\_PPO still outperforms its BIGRec-based counterpart, B\_PPO, highlighting the robustness of Flower as a reference policy.

\begin{figure}[!t]
    \setlength{\abovecaptionskip}{0.3cm}
\setlength{\belowcaptionskip}{0cm}
  \centering
  \includegraphics[width=\columnwidth]{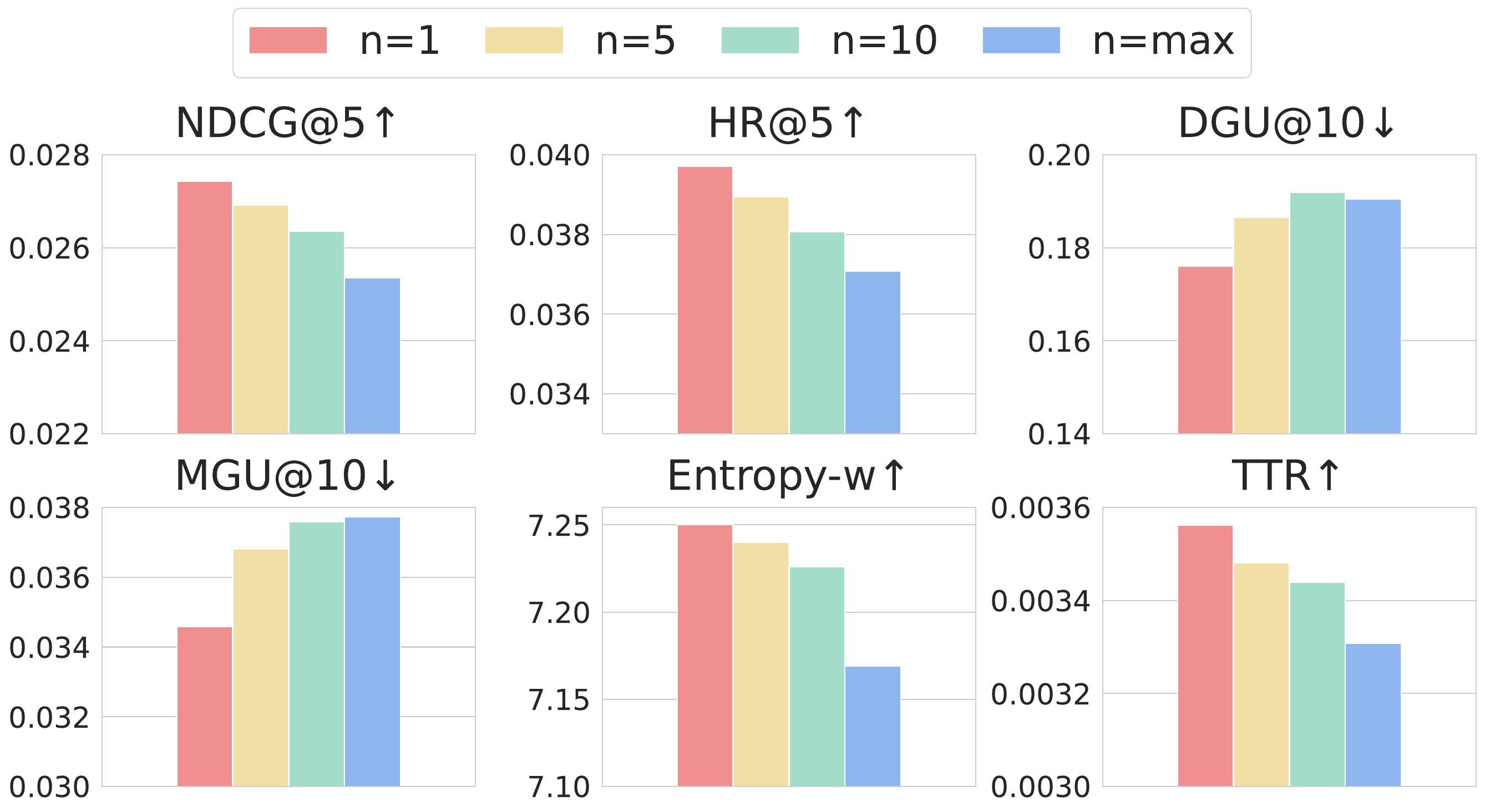} 
  % \vspace{-1mm}
  \caption{Impact of title partitioning granularities on the Video Games dataset.}  
  % \vspace{-1mm}
  \label{fig:1toN}
\end{figure}

% \subsection{Effects of Key Factors (RQ4)}
\subsection{Analysis of Key Factors in Flower (RQ4)}  
\label{sec:RQ4}  

% （这里需要用一段，来陈述为什么我们选取这三个东西来进行ablation实验），作为一个引子段落
% To comprehensively understand the contributions of Flower, we conduct ablation studies on three critical factors: reward formulation, supervision granularity, and the weighting parameter λ. These factors play distinct and important roles in shaping the model's performance: (1) reward formulation determines the target distribution used for supervision, (2) supervision granularity 通过不同的controls the strength and precision of token-level constraints, and (3) λ balances the collaboration between the flow-guided loss and the SFT loss. By systematically varying these factors, we aim to identify their individual impacts and optimize their settings to achieve a better trade-off among accuracy, fairness, and diversity.
To comprehensively evaluate the contributions of Flower, we conduct ablation studies on three critical factors: reward formulation, the granularity of title segmentation, and the hyperparameter $\lambda$ in \myeq{flower_loss}. Each factor plays a distinct and essential role in shaping the model's performance: (1) reward formulation determines the target distribution for supervision, (2) partitioning titles into subtrajectories of varying granularity enables supervision at different intensities, and (3) $\lambda$ controls the balance between the flow-guided loss and the SFT loss. By systematically varying these factors, we aim to understand their individual impacts and optimize their configurations to achieve the best trade-off among accuracy, fairness, and diversity.

\subsubsection{Effects of Reward Setting}  
\label{sec:RQ41}  
We investigate the effects of different process reward formulations. Specifically, we evaluate the results using the following three reward definitions:  
\begin{itemize}[leftmargin=*]
    \item $\log R_p$: The original process reward term, as defined in \myeq{token_reward}.
    \item $\frac{\log R_p}{p_{ui}}$: A modified process reward described in \mysec{rewards}, where the preference score $p_{ui}$ is obtained from SASRec. % and normalized to the range [1, 2].
    \item $\log(R_p \cdot p_{ui})$: Another modified process reward, with the preference score $p_{ui}$ also derived from SASRec.
\end{itemize}

The comparison of recommendation performance is shown in \mytab{reward comparision}. For \textbf{accuracy}, $\frac{\log R_p}{p_{ui}}$ achieves the best performance on the Video Games and Movies datasets. Except for NDCG on the CDs dataset, all optimal accuracy is attained by methods incorporating SASRec scores, which validates the positive impact of personalized preferences on accuracy.
For \textbf{fairness}, $\log(R_p \cdot p_{ui})$ demonstrates superior performance on the CDs and Video Games datasets compared to other methods, while $\log R_p$ performs well on only one metric of one dataset. This indicates that the effective integration of fairness and personalized preferences can simultaneously improve both accuracy and fairness. For \textbf{diversity}, $\frac{\log R_p}{p_{ui}}$ outperforms others on two datasets and ranks second only to $\log R_p$ on one dataset.
Considering all three aspects, $\frac{\log R_p}{p_{ui}}$ exhibits the best comprehensive performance.

\subsubsection{Impact of Supervision Granularity}  
\label{sec:RQ42}

Theoretically, convergence guarantees identical distributions regardless of how the states are partitioned. However, in practical scenarios, optimization is constrained to a finite number of steps, making it difficult to achieve the theoretical optimum. Therefore, ensuring sample efficiency becomes critical. To address this, we investigate how partitioning titles with varying granularities affects the effectiveness of fine-tuning.

We experiment with four granularities on the Video Games dataset: partitioning titles every 1 token (the original Flower method), 5 tokens, 10 tokens, or treating the entire title as a single state. During training, evaluations are performed every 20 steps, and the average values for each metric are calculated. As illustrated in \myfig{1toN}, accuracy, fairness, and diversity improve progressively as the granularity becomes finer. The best performance is achieved when each token is treated as an individual partition, indicating that finer-grained constraints provide stronger supervision and lead to better training outcomes.

\subsubsection{Performance Varying $\lambda$} 
\label{sec:RQ43}  

\begin{figure}[!t]
\setlength{\abovecaptionskip}{0.2cm}
  \centering
  \includegraphics[width=\columnwidth]{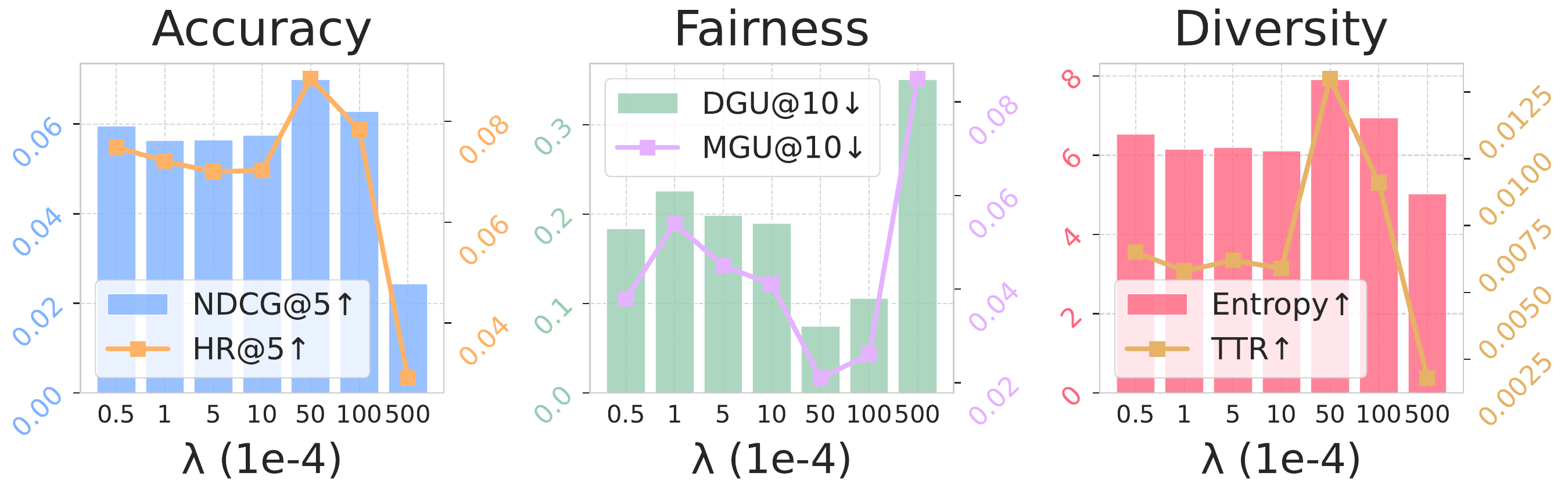} % 使用 \columnwidth 确保图片宽度适合单栏
  \vspace{-1mm}
  \caption{Performance with varying $\lambda$ on the CDs dataset.} % 添加图片标题
  \vspace{-1mm}
  \label{fig:train lamda} % 图片标签
\end{figure}

% The flow-guided loss and SFT loss can differ by 3 to 4 orders of magnitude, necessitating careful weighting for effective combination. In this subsection, we compare the model's performance with different values of $\lambda$ on the CDs dataset.

The flow-guided loss and SFT loss can differ by 3 to 4 orders of magnitude, prompting us to explore the impact of $\lambda$ across different magnitudes. We report the results on the CDs dataset.
As shown in \myfig{train lamda}, as $\lambda$ increases from left to right, the influence of the SFT loss diminishes while that of the flow-guided loss increases. Accuracy, fairness, and diversity generally exhibit a trend of first improving and then declining, with the best performance observed around $\lambda = 0.005$. When the weight of the flow-guided loss becomes excessively large, the lack of SFT loss constraints leads to a collapse in performance. Conversely, when $\lambda \leq 0.001$ (x-axis value of 10 or lower), the influence of the flow-guided loss becomes negligible, and the performance gradually converges to that of the SFT-only method.
% In \myfig{train lamda}, as $\lambda$ increases from left to right, the influence of the SFT loss diminishes while that of the flow-guided loss increases. Accuracy, fairness, and diversity generally exhibit a trend of first improving and then declining. 最佳结果在$\lambda = 0.005$附近. When $\lambda > 0.05$ (x-axis value of 500 or higher), the excessive weight of the flow-guided loss hinders the optimization of accuracy, with generated titles disproportionately favoring popular items, significantly reducing fairness and diversity. Conversely, when $\lambda \leq 0.001$ (x-axis value of 10 or lower), the effect of the flow-guided loss becomes negligible, and the performance gradually converges to that of the SFT-only method.

% \vspace{-2mm}
\section{Related Work}
\label{sec:related}

% In this section, we briefly introduce LLMs for the recommendation task and the applications of GFlowNets.

\subsection{LLMs for Recommendation}

Large Language Models (LLMs) have shown strong capabilities in text generation, reasoning, and generalization, motivating their use in personalized recommendation tasks. Supervised fine-tuning (SFT) has become a core approach for adapting LLMs to domain-specific recommendation data, significantly boosting performance \cite{bao2023bi,zhang2023collm,chen2025dlcrec}. To further align model outputs with user preferences and reduce bias, post-SFT training methods such as Direct Preference Optimization (DPO) have been proposed \cite{chen2024softmax,gao2024sprec}.

Despite these advances, SFT-based models often suffer from popularity bias, leading to filter bubbles and degraded user experience \cite{gao2023alleviating,gao2023cirs}. This is largely due to overfitting introduced by the cross-entropy loss used in fine-tuning, which biases the model toward frequently occurring items in the training set. While recent efforts represent items using unique identifier sequences \cite{Tiger,10.1145/3627673.3679569}, they remain within the CE loss framework and inherit its limitations.

In this work, we propose a novel fine-tuning paradigm that addresses the shortcomings of SFT and promotes more balanced, personalized recommendations.

\subsection{Process Supervision}
To better align LLMs with human preferences or enhance reasoning ability, a post-SFT alignment step is often introduced. Techniques like RLHF and DPO \cite{rafailov2024direct} apply outcome-level supervision by evaluating model responses holistically. However, such supervision is coarse-grained, making it hard to interpret or guide intermediate reasoning steps \cite{setlur2024rewarding}.

Recently, process supervision has gained attention for its ability to provide step-level feedback using Process Reward Models (PRMs) \cite{zhang2025lessons,li2024process}. Compared to outcome-level methods, process supervision offers more interpretable and direct optimization signals, leading to improved reasoning quality and alignment.
Most existing approaches require learning parametric PRMs---e.g., modeling step-wise correctness probabilities \cite{wang2024math,shao2024deepseekmath}---which is often impractical in recommendation tasks due to sparse user feedback. In this work, we adopt simple yet effective heuristic rewards to approximate step-level quality. This design avoids additional parameter learning and enables efficient process supervision in data-scarce settings.

\subsection{Applications of GFlowNets}

A key advantage of GFlowNets lies in their ability to sample diverse solutions while maintaining proportionality to the reward. Furthermore, GFlowNets demonstrate strong generalization capabilities, allowing them to handle states not encountered during training \cite{Bengio2021flow,Bengio2024GFlowNet,zhang2022generative,TB}. These properties make GFlowNets particularly well-suited for tasks that require exploring a wide solution space, such as molecular design \cite{jain2022biological,10.5555/3666122.3669609} and structured prediction \cite{malkin2023gflownets}. In recommendation systems, GFlowNets have also shown significant potential. For example, \citet{shuchang2023flow} use GFlowNets to introduce diversity while maintaining quality in listwise recommendations, and \citet{Liu2024retentionGflowNet} apply GFlowNets to enhance user retention while fostering exploration.

% A key advantage of GFlowNets lies in their ability to sample diverse solutions while maintaining proportionality to the reward. Additionally, GFlowNets exhibit strong generalization capabilities, enabling them to generalize to states not encountered during training \cite{Bengio2021flow,Bengio2024GFlowNet,zhang2022generative,TB}. These properties make GFlowNets particularly well-suited for tasks requiring exploration of a wide solution space, such as molecular design \cite{jain2022biological,rgfn,10.5555/3666122.3669609} and structured prediction \cite{malkin2023gflownets,deleu2022bayesian}. In recommendation systems, GFlowNets have also demonstrated promising applications. For instance, \citet{shuchang2023flow} leverage GFlowNets to introduce diversity while maintaining quality in generated listwise recommendations, and \citet{Liu2024retentionGflowNet} utilize GFlowNets to enhance user retention while encouraging exploration.

Recently, GFlowNets have been employed in fine-tuning LLMs for specific tasks. For example, \citet{hu2024amortizing} fine-tune LLMs using GFlowNets to achieve diversity in tasks such as sentence infilling, chain-of-thought reasoning, and problem-solving with external tools. \citet{Lee2025learning} adopt GFlowNets to fine-tune LLM-based attacker models, enabling the generation of diverse and effective attack prompts for Red-teaming. Similarly, \citet{yu2024flow} apply GFlowNets to train LLMs for puzzle-solving tasks, including BlocksWorld and Game24. 
In this work, we are the first to use GFlowNets to address the limitations of SFT in LLM-based next-item recommendation tasks. 
% By integrating GFlowNets, we aim to enhance diversity, fairness, and overall performance in recommendations while overcoming the biases inherent in traditional SFT approaches.

\section{Conclusion}
% This work addresses key limitations of supervised fine-tuning (SFT) in LLM-based recommendation systems, namely limited diversity and amplified popularity bias, which undermine accuracy, fairness, and personalization. These issues stem from SFT’s reliance on cross-entropy loss, leading to overfitting and reinforcement of biases in the training data.
% To tackle these challenges, we propose \textbf{Flower}, a novel fine-tuning paradigm that leverages generative flow networks (GFlowNets) for process-level supervision. Flower models the recommendation task as a flow network, where item-level rewards, derived from item frequencies in the training data, are propagated to token-level rewards for next-token prediction. This approach aligns token probabilities with reward distributions, balancing accuracy with fairness and diversity in recommendations.
% Extensive experiments on three real-world sequential recommendation datasets demonstrate Flower’s effectiveness, achieving superior results in accuracy, diversity, and fairness compared to SFT. Moreover, applying alignment methods post-Flower tuning yields better performance than applying them to conventionally fine-tuned models.

This work addresses key limitations of supervised fine-tuning (SFT) in LLM-based recommendation systems, notably limited diversity and amplified popularity bias, which hinder accuracy, fairness, and personalization. These issues largely arise from the overfitting nature of cross-entropy loss used in SFT.
To overcome these challenges, we propose \textbf{Flower}, a novel fine-tuning paradigm based on generative flow networks (GFlowNets) and process-level supervision. Flower frames recommendation as a flow network, propagating item-level rewards—derived from item frequencies—to token-level supervision. This aligns token generation with reward distributions, promoting balanced, diverse, and fair recommendations.
Experiments on three real-world sequential recommendation datasets show that Flower outperforms SFT in accuracy, diversity, and fairness. Furthermore, applying alignment methods after Flower fine-tuning yields better results than applying them on top of standard SFT models.

This study emphasizes the transformative potential of integrating diversity-seeking mechanisms into LLM-based recommendation systems. By introducing the Flow-guided fine-tuning paradigm, we address core limitations of conventional approaches, bridging the gap between accuracy, fairness, and personalization. Beyond its immediate applications, this framework lays the foundation for future innovations, including extending its applicability to diverse recommendation contexts and refining reward design to better capture nuanced user preferences and domain-specific goals.

\begin{acks}
This work is supported by the National Natural Science Foundation of China (62402470, U21B2026, 62121002, U24B20180), Anhui Provincial Natural Science Foundation (2408085QF189), the Fundamental Research Funds for the Central Universities of China (WK2100000053), and the advanced computing resources provided by the Supercomputing Center of the USTC.
\end{acks}

\bibliographystyle{ACM-Reference-Format}
\balance
\bibliography{Flower}
% \end{multicols}

\end{document}